# THE LIGHT-TIME EFFECT IN THE ALGOL TYPE ECLIPSING BINARY UZ SAGITTAE

ALEXIOS LIAKOS AND PANAGIOTIS NIARCHOS

*National and Kapodistrian University of Athens, Faculty of Physics*
*Department of Astrophysics, Astronomy and Mechanics,*
*Panepistimiopolis, GR 157 84 Zografos, Athens, Greece*
*E-mail: alliakos@phys.uoa.gr*

*Abstract*. New times of minima of the Algol-type eclipsing binary UZ Sge, obtained at the Athens University Observatory, have been used together with all reliable timings found in the literature in order to study the period variation and search for the presence of a third body in the system. Its O-C diagram is presented and apparent period changes are discussed with respect to possible Light-Time Effect (LITE) in the system. A least square method has been used to compute new light elements (updated ephemeris of binary) as well as the mass function, its minimum mass and the period of a possible third body.

## 1. INTRODUCTION

Eclipsing binaries (hereafter EBs) often display changes of their orbital period. The light-time effect (hereafter LITE) in eclipsing binaries produces period variations measurable by their times of minima. In some cases, the reason for these periodic changes can be the presence of a third body in the system. The eclipsing pair, orbiting around a common center of mass, changes its distance from us periodically and the times of minima change too. LITE was first discussed by Irwin (1959), and the necessary criteria have been mentioned by Frieboes-Conde and Herczeg (1973) and they are the following: (i) agreement between the O-C data points and the theoretical LITE curve, (ii) primary and secondary minima show the same displacement in time**,** (iii) reasonable mass function of the third body and (iv) variations in the radial velocities of the components of the eclipsing pair in systematic and long-term (several years) spectroscopic observations. The criteria (i)-(iii) are satisfied in the system studied in this paper.

UZ Sge (AN 435.1936 = GSC 1626-1289, $\alpha_{2000}$ = $20^h12^m16.21^s$, $\delta_{2000}$ = +19° 20´55.5´´) was discovered to be a variable by Guthnick and Schneller (1939). Using the method of photometric parallaxes, Bancewicz & Dworak (1980) calculated the absolute parameters of the system and its spectral type to be A0. The first spectroscopic observation was performed by Halbedel (1984) who determined the spectral type of the system as A0. Unfortunately, neither spectroscopic nor photometric analysis has been performed so far.

## 2. OBSERVATIONS

The system was observed with the 40-cm Cassegrain telescope of the Observatory of the University of Athens, equipped with the ST-8XMEI CCD camera using the Bessell R-filter. The observations were carried out during two nights in July 2007 and in one night in August 2007. Differential photometry was used in order to measure the light variation. The star GSC 1626-1303 was selected as the comparison star and its constancy was checked by the star GSC 1626-1845. Two secondary and one primary time of minima were obtained and they are given in Table 1.

*Table 1*

The times of minima obtained from our observations

| HJD( 2400000.0+) | Type | Error |
|---|---|---|
| 54293.45120 | II | 0.00023 |
| 54313.39219 | II | 0.00021 |
| 54314.49758 | I | 0.00013 |

## 3. O-C DIAGRAM ANALYSIS

In order to analyze the O-C diagram of the system the least squares method has been used. The computation of the third body's orbit is an inverse problem with 5 parameters to be found. These are: $P_3$, the orbital period of the third body around the barycenter of the triple system; $T_0$, the Julian Date of the periastron passage of the third body; $A$, the semi-amplitude of O-C; $\omega$, the argument of the periastron; and $e$, the eccentricity of the third body's orbit. The epoch of a primary minimum $JD_0$ in the ephemeris of the eclipsing binary, and the period $P$ are also included as independent parameters in the calculation. The assigned weight (w) for the individual observations was based on the reliability of each data point. We used w =1 for the visual points, 5 for the photographic data and 10 for photoelectric and CCD observations.

In the present analysis, 80 times of minima taken from Kreiner's database (private communication) and 3 times of minima derived from our CCD observations were used. The following ephemeris (Kreiner et al. 2001) was used to compute, initially, the O-C residuals of all compiled times of minima:

$$T = 2445861.4115 + 2.21574259 \times E \tag{1}$$

A sinusoidal fit in the O-C diagram is shown in Fig. 1, while the (O-C) residuals for all times of minima are shown in Fig. 2.

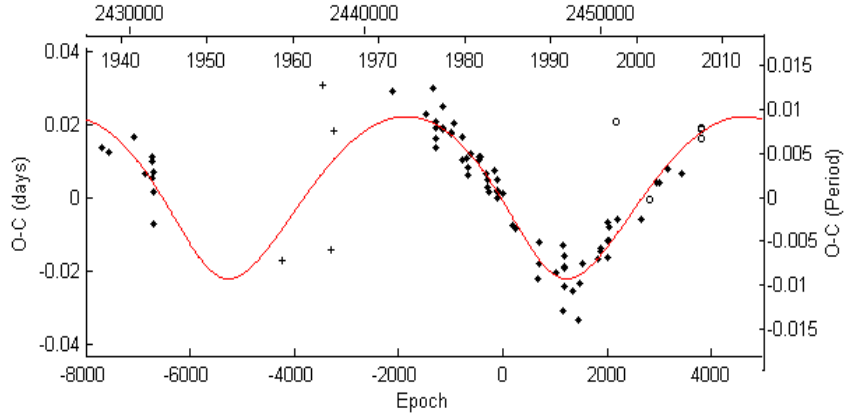

Fig. 1. The O-C diagram of UZ Sge fitted by a sinusoidal curve. The full rhombs represent the visual data, the crosses the photographic data and the open circles the photoelectric and CCD observations.

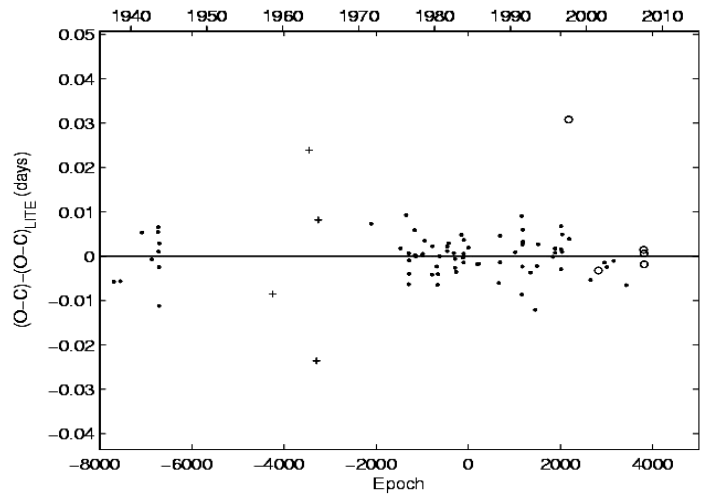

Fig. 2. The O-C diagram after subtraction of LITE solution.
(The meaning of symbols is the same as in Fig. 1.)

In order to derive the minimal mass (for i = 90°) of the third body, we used the masses $M_1 = 2.89\ M_\odot$ and $M_2 = 1.97\ M_\odot$ (Brancewicz & Dworak, 1980) for the masses of the two components of the EB, respectively. The orbital parameters of the third body, its mass function and minimal mass are listed in Table 2.

*Table 2*

Results for the system UZ Sge

| Parameter | Value | Parameter | Value |
|---|---|---|---|
| $JD_0$ (HJD) | 2445861.41303 ± 0.00178 | $T_0$ (HJD) | 2448084.094 ± 912.782 |
| P (days) | 2.2157454 ± 0.0000006 | ω (deg) | 251.9 ± 23.9 |
| $P_3$ (yrs) | 39.36 ± 0.69 | $f(m_3)$ | 0.03654 ± 0.00002 |
| A (days) | 0.022 ± 0.001 | $M_{3,\ min}\ (M_\odot)$ | 0.6955 ± 0.0004 |
| e | 0.27 ± 0.06 | $\chi^2$ | 0.0087917931157 |

## 4. SUMMARY AND CONCLUSIONS

New LITE parameters and ephemeris ($JD_0$, P) of UZ Sge have been derived by means of an O-C analysis. The observational data and the derived values of the LITE parameters confirm the fulfillment of the first three necessary criteria for LITE, mentioned in section 1. An inspection of Fig. 2 reveals that the residuals show no more periodicities that would allow us to search for additional effects. Regrettably, no complete light curve analysis has been obtained for this system so far, so the existence of the predicted third body cannot be confirmed yet. A complete analysis using other methods, e.g. radial velocity measurements and/or photometric light curve analysis, is needed.

Acknowledgments. We thank P. Zasche for providing the matlab code to compute the LITE elements and Prof. J. Kreiner for making available the list with the times of minima. This work has been financially supported by the Greek-Czech project of collaboration GSRT 214-γ of the Ministry of Development (Special account for Research Grants 70/3/8680 of the National & Kapodistrian University of Athens, Greece), and by UNESCO-BRESCE.

REFERENCES

Brancewicz, H.K., Dworak, T.Z.: 1980, *Acta Astronomica*, **30**, 4
Frieboes-Conde, H., Herczeg, T.: 1973, *A&AS*, **12**, 1
Halbedel, E.M.: 1984, *IBVS*, **2549**, 1
Irwin, B.J.: 1959, *AJ*, **64**, 149
Kreiner, J.M., Kim, C.H., Nha, S.: 2001, An Atlas of O-C diagrams OF Eclipsing Binary Stars, Cracow, Poland